\documentclass[12pt,preprint]{aastex}

\newcommand{\etal}{et~al.\ }

\newcommand{\cmsq}{\hbox{cm$^{-2}$}}

\newcommand{\nh}{\hbox{${N}_{\rm H}$}}

\newcommand{\chandra}{{\emph{Chandra}}}
\newcommand{\cxo}{{\emph{Chandra X-ray Observatory}}}
\newcommand{\xmm}{\emph{XMM-Newton}}
\newcommand{\asca}{{\emph{ASCA}}}
\newcommand{\rosat}{\emph{ROSAT}}
\newcommand{\hst}{\emph{HST}}

\newcommand{\mg} {MG~0414+0534}
\newcommand{\he} {HE~1104$-$1805}
\newcommand{\pks} {PKS~1830$-$211}

\begin{document}

\def\sarc{$^{\prime\prime}\!\!.$}
\def\arcsec{$^{\prime\prime}$}
\def\arcmin{$^{\prime}$}
\def\degr{$^{\circ}$}
\def\seco{$^{\rm s}\!\!.$}
\def\ls{\lower 2pt \hbox{$\;\scriptscriptstyle \buildrel<\over\sim\;$}} 
\def\gs{\lower 2pt \hbox{$\;\scriptscriptstyle \buildrel>\over\sim\;$}} 
 
\title{Probing The Dust-To-Gas Ratio of $z > 0$ Galaxies Through Gravitational Lenses}

\author{Xinyu Dai\altaffilmark{1}, Christopher S. Kochanek\altaffilmark{1}, George Chartas\altaffilmark{2}, and Smita Mathur\altaffilmark{1}}

\altaffiltext{1}{Department of Astronomy,
The Ohio State University, Columbus, OH 43210,
xinyu@astronomy.ohio-state.edu, ckochanek@astronomy.ohio-state.edu, smita@astronomy.ohio-state.edu}
\altaffiltext{2}{Department of Astronomy and Astrophysics,
The Pennsylvania State University, University Park, PA 16802,
chartas@astro.psu.edu}

\begin{abstract}
We report the detection of differential gas column densities in three gravitational lenses, \mg, \he, and \pks.  
Combined with the previous differential column density measurements in B~1600+434 and Q~2237+0305 and the differential 
extinction measurements of these lenses, we probe the dust-to-gas ratio of a small sample of cosmologically distant normal
galaxies.  We obtain an average dust-to-gas ratio of $E(B-V)/$\nh $=(1.4\pm0.5)\times$ 10$^{-22}$ mag cm$^2$ atoms$^{-1}$ 
with an estimated intrinsic dispersion in the ratio of $\simeq 40$\%.  This average dust-to-gas ratio is consistent 
with the average Galactic value of $1.7 \times 10^{-22}$ mag cm$^2$ atoms$^{-1}$ and the estimated intrinsic
dispersion is also consistent with the 30\% observed in the Galaxy.  
\end{abstract}

\keywords{dust-to-gas ratio}

\section{Introduction}
Gravitational lenses are valuable tools in many astrophysical applications.  One of them is to study the properties of the interstellar medium (ISM) 
in the lens galaxies at the locations where we observe the quasar images.
By studying the \emph{differences} in the absorption (or extinction) between multiple images, any contamination from the Galaxy in
the foreground or the quasar host galaxy in the background is essentially eliminated and we can be confident that we are probing the ISM of the lens galaxy.
Several groups \citep[e.g.,][]{na91,fal99,thb00,mo02,wu03,mu04,me05} have studied dust extinction and the dust extinction law in lens galaxies.
In X-rays, \citet{da03} and \citet{dk05} measured the differential column densities in two lenses Q~2237+0305 and B~1600+434 using a similar 
method.  
By combining the X-ray and optical measurements of the differences in gas and dust properties between the lensed images, we can estimate the 
dust-to-gas ratio $\Delta E(B-V)/\Delta\nh$ of the lens galaxies
assuming the differences between the extinction and gas absorption are due to the same parcel of the ISM.  

The dust-to-gas ratio, normally calculated from $E(B-V)/\nh$, is a basic property of the ISM, but our direct knowledge of dust-to-gas ratios is largely based 
on our own Galaxy.  In particular, \citet{bsd78} measured the dust-to-gas ratios of the ISM towards 100 stars in the Milky Way and found a 
correlation of $E(B-V)/\nh = 1.7\times10^{-22}$ mag cm$^{2}$ atoms$^{-1}$ with typical scatter of $\sim$30\%.
Subsequent studies have extended these measurements to additional lines of sight in our Galaxy and to the LMC
and SMC, but the basic results are little altered aside from some evidence that the dust-to-gas ratios of the 
LMC and SMC are modestly lower ($\sim 30$\%) than in the Galaxy (see the review by Draine 2003 and references therein). 
However, because of the difficulties in determining the absorption column densities and the extinction of distant galaxies, 
little is known about the dust-to-gas ratios of their ISM.  Just as with studies of extinction laws, gravitational lenses provide the 
first potential probe for studying the dust-to-gas ratios of cosmologically distant galaxies ($0<z<1$) with the precision of studies
of the ISM in the Galaxy.  Here we measure the differential column density of three additional lenses and combine them with existing 
measurements for two lenses to estimate the average dust-to-gas ratio of the lens galaxies.

\section{Sample Selection and Data Reduction}
Over twenty gravitational lenses have been observed in X-rays with the \cxo\ \citep{we02}.
From these systems, we selected the seven systems from the \citet{fal99} survey of extinction in lens galaxies
with differential extinction measurements exceeding $\Delta E(B-V) > 0.05$ mag. 
For two of these systems, LBQS~1009$-$0252 and MG~2016+112, the \chandra\ spectra of the individual images contain too few photons to 
allow an analysis of differential absorption, and the differential absorptions in B~1600+434 and Q~2237+0305 have been reported 
previously \citep{da03,dk05}.  Here we analyze the remaining three lenses, \mg, \he, and \pks.

\mg\ and \he\ were observed with the Advanced CCD Imaging Spectrometer \citep[ACIS,][]{g03} on board \chandra\ for 97 and 49 ks on January 8, 2002 and June 10, 2000, respectively.  \pks\ was observed with the High Energy Transmission Grating Spectrometer \citep[HETGS,][]{ca05} and ACIS on board \chandra\ for 51 ks on June 25, 2001.  The \chandra\ data were reduced with the \verb+CIAO+ software tools provided by the \chandra\ X-ray Center (CXC) following the standard threads on the CXC website.\footnote{The CXC website is at http://cxc.harvard.edu/.}  Only events with standard \asca\ grades of 0, 2, 3, 4, and 6 were used in the analysis.  We improved the image quality of the data by removing the pixel
randomization applied to the event positions by the standard pipeline.
In addition, we applied a subpixel resolution technique \citep{t01,m01} to the events on the S3 chip of ACIS where the lensed images are located.

\section{Spectral Analysis}
We extracted spectra for each of the lensed quasar images.  Although the \pks\ observation was a grating observation, we simply analyzed the spectra obtained from the zeroth order image to extract and study spectra of the two quasar images separately.
We did not analyze the grating spectrum of \pks\ because it cannot resolve the spectra of the two images, which is essential in this study.
We fitted the spectra of the lensed quasars using \verb+XSPEC V11.3.1+ \citep{a96} over the 0.3--8 keV observed energy range.
The X-ray spectrum of the $i$th image can be generally modeled as
\begin{eqnarray}
N_i(E, t) & = & N_{0, i}\left(\frac{E}{E_0}\right)^{-\Gamma(t-\Delta t_i)} \nonumber\\
& & \exp\left\{-\sigma(E)\nh_{,Gal}-\sigma\left[E(1+z_l)\right]\nh_{,i}-\sigma\left[E(1+z_s)\right]\nh_{,Src}(t-\Delta t_i)\right\},
\end{eqnarray}
where $N_i(E, t)$ is the number of photons per unit energy interval, $\nh_{, Gal}$, $\nh_{,i}$, and $\nh_{,Src}$ are the equivalent hydrogen column density at our Galaxy, the lens galaxy at image $i$, and the source galaxy, $\sigma(E)$ is the photo-electric absorption cross-section, and $\Delta t_i$ is the time-delay. 
The difference between the absorption at the lens, $\Delta\nh$, can be obtained by analyzing the individual spectra of the lensed images.
We neglected the time-delay between the images and the absorption at the source redshift in this analysis.
We used the standard \verb+wabs+ and \verb+zwabs+ models in \verb+XSPEC+ to model the Galactic absorption and absorption at the lens redshift.  The \verb+wabs+ and \verb+zwabs+ models use cross sections from \citet{mm83} and assume a solar elemental abundance from \citet{ae82}.

\subsection{\mg}
\mg, discovered by \citet{he92}, is a four-image lens system with a source redshift of $z_s = 2.639$ \citep{la95} and a lens redshift of $z_l = 0.9584$ \citep{tk99}.  The \chandra\ image of \mg\ was presented in \citet{ch02}.  We extracted the spectra of the relatively well separated images A (A1 and A2 combined), B and C from the 97 ks \chandra\ observation.  The A1 and A2 images are too close (0\sarc4 apart) to obtain the individual spectra of the images.

\begin{figure}
\epsscale{0.8}
\plotone{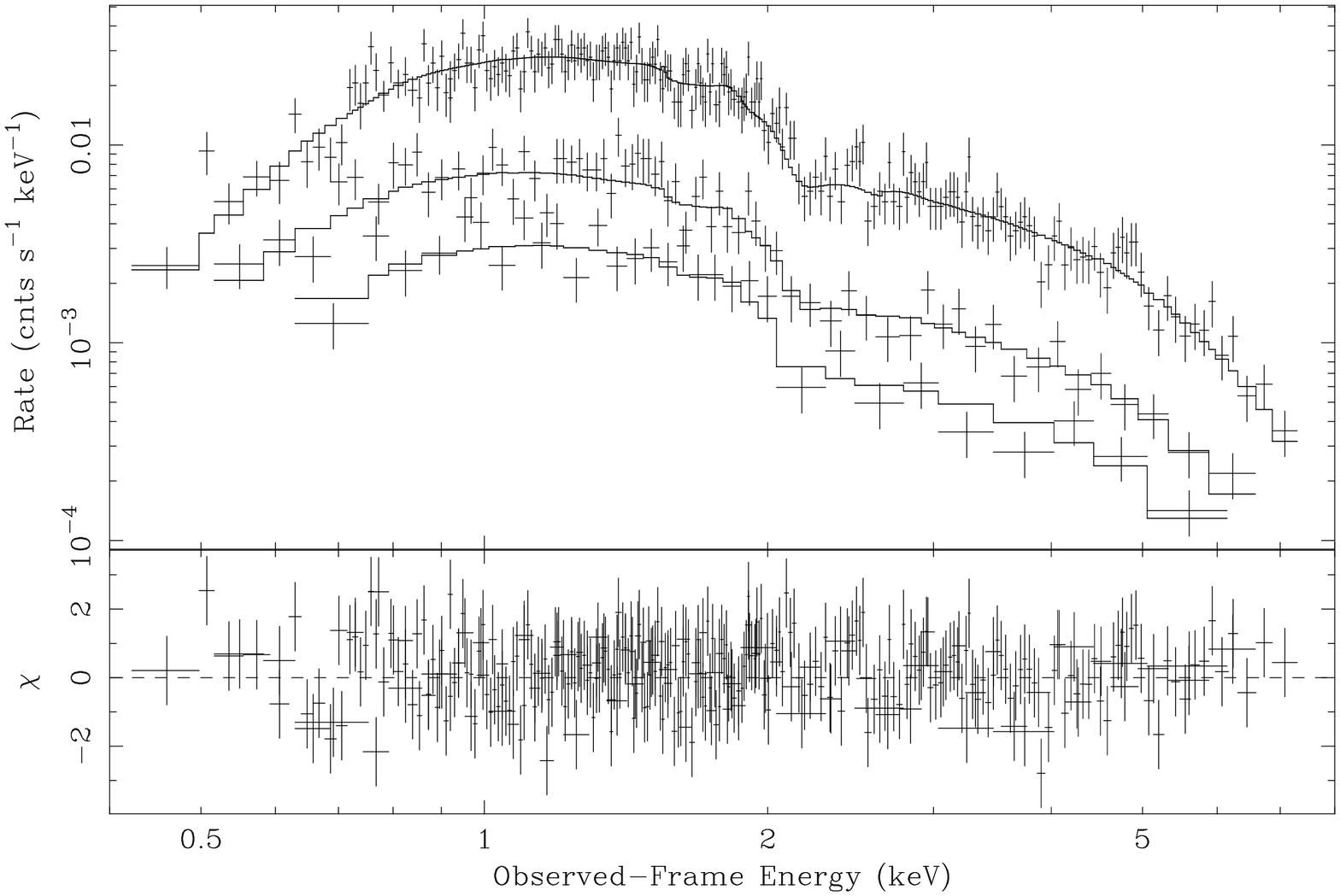}
\caption{Spectra of images A (top), B and C (bottom) of \mg. \label{fig:mg}}
\end{figure}

We modeled the spectra with a power-law modified by absorption at the lens and Galactic absorption.  The Galactic absorption was fixed as 
$\nh\ = 0.11\times10^{22}$~\cmsq\ \citep{d90}.  We fitted the three spectra simultaneously using the same power-law photon index for all spectra but 
differing amounts of absorption at the lens redshift for each image.  The results are presented in Table~\ref{tab:spec}.  Our estimate of the power-law 
photon index, $\Gamma=1.69^{+0.04}_{-0.01}$, is consistent with the range of values from \citet{ch02}.  Figure~\ref{fig:mg} shows the spectra of the three images and their best fit models.
It is clear from the spectra that the spectrum of image A is more heavily absorbed than that of images B and C.  We estimate that the differences in the 
column densities are $\Delta\nh_{A,B} = (0.33\pm0.10) \times 10^{22}$~\cmsq, $\Delta\nh_{A,C} = (0.20\pm0.15) \times 10^{22}$~\cmsq and 
$\Delta\nh_{C,B} = 0.1^{+0.2}_{-0.1} \times 10^{22}$~\cmsq.  We used \verb+XSPEC+ to explore the full parameter space of the column
densities in order to correctly estimate the uncertainties in the differential column densities.  We also verified that adding 
absorption in the source does not alter the estimates of the differential column densities between the images in the lens galaxy.

\begin{deluxetable}{cccccccc}
\tabletypesize{\scriptsize}
\tablecolumns{8}
\tablewidth{0pt}
\tablecaption{Spectral Fitting Results \label{tab:spec}}
\tablehead{
\colhead{} &
\colhead{} &
\colhead{Galactic \nh} &
\colhead{\nh\ (Image A)} &
\colhead{\nh\ (Image B)} &
\colhead{\nh\ (Image C)} &
\colhead{} &
\colhead{}
\\
\colhead{Quasar} &
\colhead{Model\tablenotemark{a}} &
\colhead{($10^{22}$~\cmsq)} &
\colhead{($10^{22}$~\cmsq)} &
\colhead{($10^{22}$~\cmsq)} &
\colhead{($10^{22}$~\cmsq)} &
\colhead{$\Gamma$} &
\colhead{$\chi^{2}_{\nu}(dof)$} 
}

\startdata
\mg\ & wabs(zwabs(pow)) & 0.11 (fixed) & $0.86\pm0.06$ & $0.53\pm0.10$ & $0.64^{+0.22}_{-0.13}$ & $1.69_{-0.01}^{+0.04}$ & 0.95(318) \\
\he\ & wabs(zwabs(pow)) & $<0.004$ & $0.06\pm0.03$ & $<0.01$ & \nodata & $1.68\pm0.06$ & 0.88(84) \\ 
\pks\ & wabs(zwabs(pow)) & 0.22 (fixed) & $1.0^{+0.4}_{-0.3}$ & $2.8\pm0.6$ & \nodata & $1.02\pm0.06$ & 1.08(237) \\
\enddata

\tablecomments{Simultaneous fits to the individual spectra of the lensed quasars.  The spectra are constrained to have the same intrinsic power-law photon index but can have different absorption column densities at the redshift of the lens.}
\tablenotetext{a} {The wabs, zwabs, and pow represent the Galactic absorption, absorption at the lens, and the intrinsic power-law spectrum, respectively.}
\end{deluxetable}

\citet{fal99} measured the differential extinction of the four images as $\Delta E(B-V) =$ 0.09, 0.31, 0.02 and 0 mag for images A1, A2, B and C, 
respectively.  The differential extinction between images B and C is very small.  The differential extinction between image A1 and A2 is large with 
$\Delta E(B-V)=0.22$~mag.  We used the two brightest images to estimate the dust-to-gas ratio.  Because we only analyzed the X-ray spectrum of the 
image A (A1 and A2 combined), we used an average differential extinction of $\Delta E(B-V)=0.18$~mag between images A and B in our dust-to-gas ratio 
estimate.  Combined with our differential absorption measurement, we obtained a dust-to-gas ratio of 
$E(B-V)/\nh = (0.55\pm0.33)\times10^{-22}$ mag cm$^2$ atoms$^{-1}$.  The error-bar was estimated by replacing the average differential extinction 
with values for the A1 and A2 images.  This ambiguity dominates any interpretation of the differential absorption measurement.

\subsection{\he}
\begin{figure}
\epsscale{0.8}
\plotone{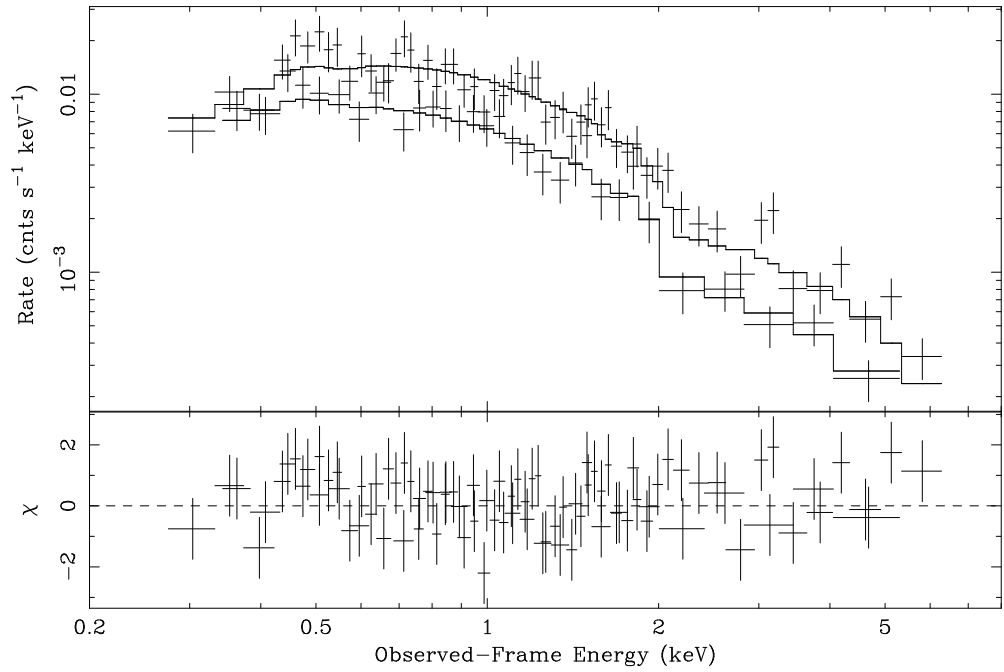}
\caption{Spectra of images A (top) and B (bottom) of \he. \label{fig:he}}
\end{figure}

\he\ is a two-image lens discovered by \citet{wi93} with a source redshift of $z_s = 2.319$ \citep{wi93} and a lens redshift of $z_l = 0.729$ \citep{li00}. 
The two images are well separated ($\Delta\theta = $ 3\sarc2), so we can easily extract the individual spectra of both images.  
The combined spectrum of both images of \he\ was previously analyzed by \citet{da04}.
In fitting the combined spectrum, as done by \citet{da04}, the low total absorption obtained indicates
that adding a small component of Galactic absorption ($\nh = 0.05\times10^{22}$~\cmsq) suffices to fit the data with a power law index of $\Gamma=1.86^{+0.06}_{-0.04}$ for the intrinsic spectrum.   
We again see the advantages of using differential measurements to
eliminate ambiguities as to the location of the absorption.
We obtained a differential column density between the two images of $\Delta\nh_{A,B} = (0.055\pm0.030) \times 10^{22}$~\cmsq\ (Table~\ref{tab:spec}) by placing the absorption at the lens redshift.  
Figure~\ref{fig:he} shows the spectra of the two images and their best fit models.  The 
power-law photon index, $\Gamma=1.68\pm0.06$, obtained in this model is slightly harder than the values from \citet{da04}.    
Curiously, we find that image A shows more absorption than image B, 
which is the reverse of the differential extinction estimated by \citet{fal99}.  This is probably due to chromatic microlensing of
the images by the stars in the lens galaxy being misinterpreted as extinction, but it could also be explained by giving the ISM
very different dust-to-gas ratios at the positions of the two images.  
Recent observations of \he\ (Wisotzki et al. 1993, 1995; G\'omez-\'Alvarez et al. 2004) indicates that the chromaticity disappears when the microlensing perturbations decrease, supporting the chromatic microlensing interpretation. 
In addition, it is possible that the amount of ISM varied between the optical and X-ray observations.  However, considering the short time-scale, it is unlikely that the ISM of a normal galaxy would vary significantly.
In any case, we will be unable to include this lens
in an estimate of the mean dust-to-gas ratio.

\begin{figure}
\epsscale{0.8}
\plotone{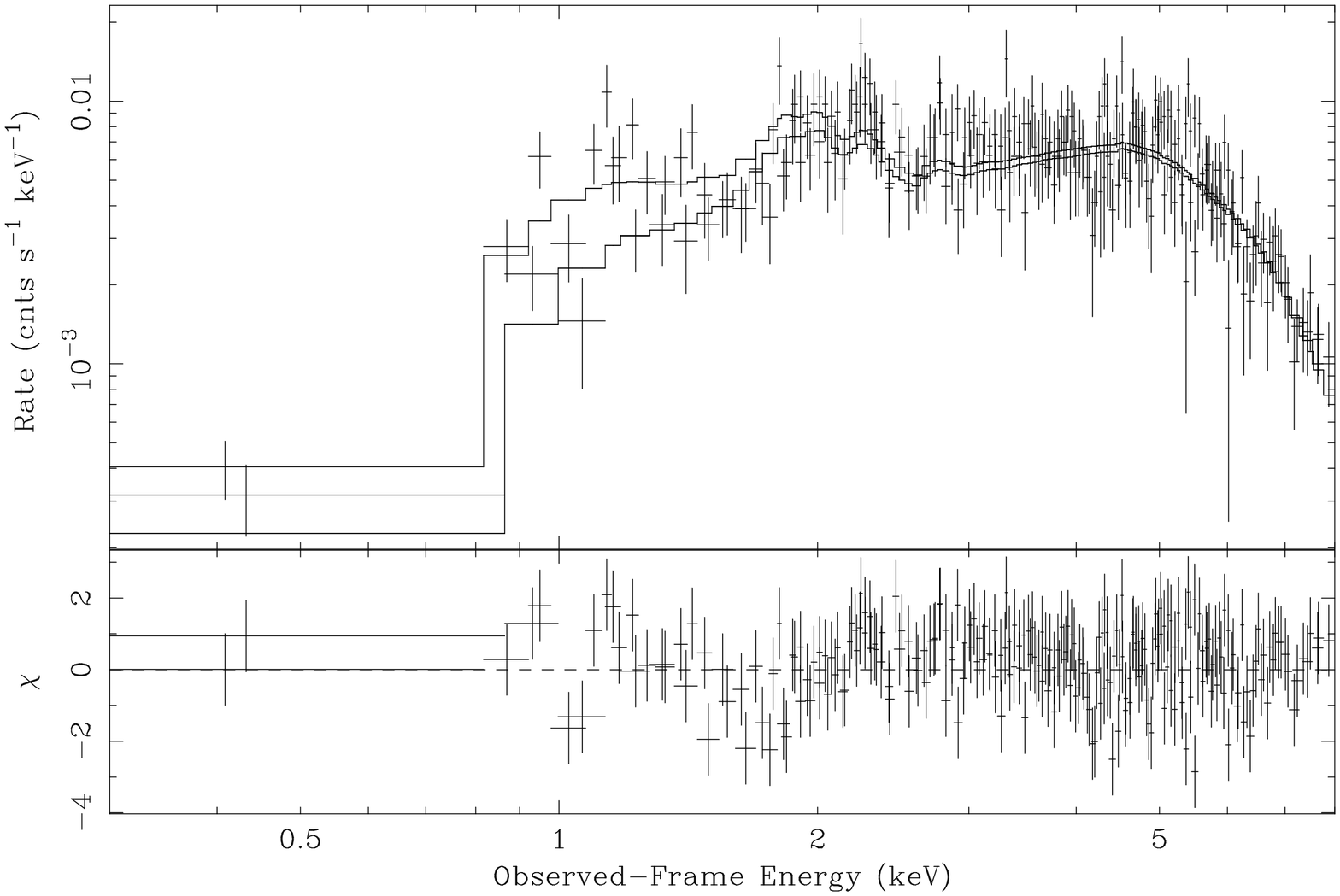}
\caption{Spectra of images A (top) and B (bottom) of \pks. \label{fig:pks}}
\end{figure}

\subsection{\pks}
\pks\ \citep{ps88,su90,ja91} consists of two $z_s =2.507$ \citep{li99} quasar images separated by 1\arcsec\ and lensed by a  $z_l = 0.886$ \citep{wc96,ge97} 
spiral galaxy \citep{wi02}.  Using our standard method, but applying it to the zeroth order HETGS image, we find that $
\Delta\nh_{B,A} = 1.8^{+0.5}_{-0.6}\times 10^{22}$~\cmsq\ (Table~\ref{tab:spec}) with the Galactic absorption fixed at $\nh = 0.22\times10^{22}$~\cmsq\ \citep{d90}.  Figure~\ref{fig:pks} shows the spectra of the two images and their best fit models.
We obtained an extremely hard power-law photon index, $\Gamma=1.02\pm0.06$, considerably harder than the values obtained from most other radio-loud quasars \citep[$\langle \Gamma \rangle \sim$ 1.5--1.6, e.g.,][]{ca97,rt00} or blazars \citep[$\langle \Gamma \rangle \sim$ 2.2, e.g.,][]{do01}.
However, it is consistent with the previous \rosat\ result \citep[$\Gamma=1^{+2}_{-1}$,][]{mn97}.

We also estimated the absorption from the combined spectrum of the two images to compare with previous measurements.
Again, we fixed the Galactic absorption at $\nh = 0.22\times10^{22}$~\cmsq.
The best fit to the combined spectrum of 
$\nh=(1.8\pm0.4) \times 10^{22}$~\cmsq\ coincidentally matches
the differential absorption measurement.
The combined absorption and the power-law photon index obtained, $\Gamma=1.01\pm0.06$, also agree with the analysis of the combined spectrum by \citet{de05}.
This combined absorption is somewhat lower than the estimates from molecular absorption in the radio ($N(H_2)\simeq 2.5 \times 10^{22}$~\cmsq, Wiklind \& Combes 1996,
1998, Gerin et al. 1997) or the column of $3.5 \times 10^{22}$~\cmsq\ required to match combined \rosat\ spectrum of the two
images with an assumed X-ray photon index of $\Gamma=2.2$ (Mathur \& Nair 1997).  However, it is consistent with their measurement of $\nh=1.7^{+2.9}_{-1.5} \times 10^{22}$~\cmsq\ allowing for a harder power-law photon index of $\Gamma=1^{+2}_{-1}$ \citep{mn97}.
The very large absorption difference, $\Delta\nh_{B,A} = 1.8^{+0.5}_{-0.6}\times 10^{22}$~\cmsq, is consistent with the 
large differential extinction of $\Delta E(B-V)=3.00$~mag \citep{fal99} between images B and A.  Combining these measurements, the dust-to-gas 
ratio is $E(B-V)/\nh = (1.7\pm0.6)\times10^{-22}$ mag cm$^2$ atoms$^{-1}$, consistent with the standard Galactic value.  

\section{Discussion}
\begin{deluxetable}{ccccccc}
\tabletypesize{\scriptsize}
\tablecolumns{7}
\tablewidth{0pt}
\tablecaption{The Dust-To-Gas Ratio of High Redshift ($z>0$) Galaxies \label{tab:dtg}}
\tablehead{
\colhead{} &
\colhead{} &
\colhead{} &
\colhead{} &
\colhead{$\Delta \nh$} &
\colhead{$\Delta E(B-V)$} &
\colhead{$E(B-V)/\nh$} 
\\
\colhead{Lens} &
\colhead{$z_l$} &
\colhead{type} &
\colhead{Between Images} &
\colhead{($10^{22}$~\cmsq)} &
\colhead{(mag)} &
\colhead{($10^{-22}$ mag cm$^2$ atoms$^{-1}$)} 
}
\startdata
\mg       & 0.9584 & elliptical & A, B & $0.33\pm0.10$          & $0.18$\tablenotemark{a} & $0.55\pm0.33$ \\   
\he        & 0.729 & elliptical & A, B & $0.055\pm0.030$        & $-0.07\pm0.01$          & \nodata \tablenotemark{b} \\
B~1600+434  & 0.41   & spiral & B, A & $0.26^{+0.17}_{-0.12}$ & \phs$0.10\pm0.03$           & \phd$0.38\pm0.25$ \tablenotemark{c} \\
\pks       & 0.886  & spiral & B, A & $1.8^{+0.5}_{-0.6}$    & \phs$3.00\pm0.13$           & $1.7\pm0.6$ \\
Q~2237+0305 & 0.0395 & spiral\tablenotemark{d} & A, C & $0.04\pm0.03$          & \phs$0.11\pm0.03$           & $2.8\pm2.2$ \\
\enddata

\tablenotetext{a} {The average extinction value of images A1 and A2 was used in the calculation and the error-bar is estimated by replacing the average differential extinction with values for the A1 and A2 images.}
\tablenotetext{b} {The dust-to-gas ratio was not estimated for \he\ because the the differential extinction and \nh\ column measurements give opposite signs.}
\tablenotetext{c} {The number provided in \citet{dk05} is the gas-to-dust ratio rather than the dust-to-gas ratio.}
\tablenotetext{d} {Q~2237+0305 is lensed by the central bulge of a spiral galaxy.}
\end{deluxetable}

\begin{figure}
\epsscale{0.8}
\plotone{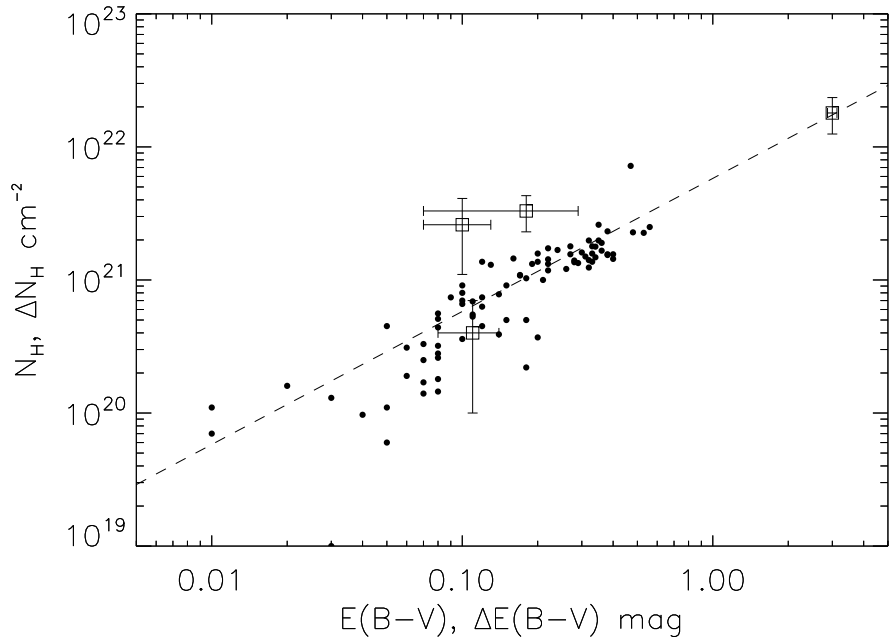}
\caption{$\Delta \nh$ versus $\Delta E(B-V)$ for the gravitational lenses (squares) and \nh\ versus $E(B-V)$ for Galactic stars from Bohlin \etal (1978, filled circles).  The dashed line shows the mean Galactic relation. \label{fig:dtg}}
\end{figure}

Table~\ref{tab:dtg} summarizes the five systems for which we have estimates of both the differential extinction and the absorption 
column density.  This includes the three lenses analyzed here (\mg, \he\ and \pks) and the previously analyzed 
systems Q~2237+0305 \citep{da03} and B~1600+434 \citep{dk05}.  We did not include B~0218+357, as the dust-to-gas ratio 
estimated by \citet{fal99} was based on very uncertain estimates ($N(H_2)\sim (0.2-5.0)\times 10^{23}$~cm$^{-2}$) of
the gas column density from molecular absorption measurements in the radio \citep{wc95,ge97,cry93}.  We considered 
only the two brightest images in each system, and for Q~2237+0305, where several extinction measurements are 
available, we used the value from \citet{fal99} to be consistent with other lenses.  The values for Q~2237+0305 are consistent with those measured by \citet{ag00}.  For estimating the average 
dust-to-gas ratio we must exclude \he\ (see \S3.2).  If we fit the four remaining lenses for the mean gas-to-dust 
ratio including the uncertainties in both the absorption and the extinction,
as well as allowing for additional intrinsic scatter in the ratio beyond the measurement errors, we find
a mean ratio of $(1.4\pm0.5) \times 10^{-22}$ mag cm$^2$ atoms$^{-1}$ and estimate that there is an intrinsic
scatter in the ratio of $\simeq 40\%$ beyond our measurement errors.  If we allow for no intrinsic scatter we
find a ratio of $(1.5\pm0.3) \times 10^{-22}$ mag cm$^2$ atoms$^{-1}$.   These estimates are consistent with the 
average Galactic value of $1.7\times 10^{-22}$ mag cm$^2$ atoms$^{-1}$ but show a modestly larger
intrinsic scatter than the locally observed 30\% \citep{bsd78}.
The estimates of the intrinsic scatter is consistent with local estimates given our small sample size.
Figure~\ref{fig:dtg}
compares the results for the lens galaxies to the local stellar sample from \citet{bsd78}.  

The lens galaxies of our small sample consist of three spiral and one elliptical galaxies, although the quasar in Q~2237+0305 
is lensed by the central bulge of the spiral galaxy (Table~\ref{tab:dtg}).  Considering the environmental and redshift 
differences, it is interesting that we obtained dust-to-gas ratios consistent with those of our Galaxy. It suggests
that the dust-to-gas ratio of the ISM is similar in all normal galaxies.  It may not, however, be universal, since 
\citet{ma01} found that the circumnuclear regions of local AGN may have a very low dust-to-gas ratio 
(a factor of $\sim$3--100 lower than Galactic), possibly because of grain formation and destruction processes
peculiar to the local environment created by the AGN \citep{ma01,mmo01}.
Recently, models of evolution of dust-to-gas ratios are discussed in several papers \citep[e.g.,][]{dw98,ed01,in03}.  However, we are unable to compare the details of the models with our small sample of data.

It is relatively straight forward to improve on these results by systematically obtaining resolved X-ray spectra of 
lenses showing significant differential extinctions.  Like most X-ray spectral studies, the required integration
times are fairly long ($\sim 50$~ksec).  
The differential absorption measurement requires resolved spectra from individual lensed images.
This means that
\xmm\ cannot be used despite its higher sensitivity -- the high spatial resolution of \chandra\ is required.
For systems with modest differential extinctions ($\Delta E(B-V) < 0.1$~mag) there are probably significant systematic
uncertainties created by chromatic microlensing where color differences created by microlensing are misinterpreted
as extinction \citep[e.g.,][]{wu03}.  
The X-ray emitting region is probably not as spatially extended as the optical
accretion disk, so that the entire X-ray continuum is magnified by roughly a constant at a given epoch, unless the observations occur during a high microlensing magnification event.  Thus, the column density estimates obtained from the absorption in the soft X-ray spectra should be insensitive to microlensing effects.  
Focusing on 
lenses with significant differential extinction will also minimize the possibility that significant differences
in the properties of the ISM between lensed images will skew the results either because the extinction laws
are very different \citep[e.g.,][]{mu04,mc05} or because the dust-to-gas ratios are
very different.   With a large enough sample, it would be possible to study the evolution of the ISM with 
redshift.

\acknowledgements
We gratefully acknowledge the financial support by \hst\ grant GO-9375 and CXC grant GO3-4154X.


\clearpage

\begin{thebibliography}

\bibitem[Agol, Jones \& Blaes(2000)]{ag00} Agol, E., Jones, B., \& Blaes, O. 2000, \apj, 545, 657

\bibitem[Anders \& Ebihara (1982)]{ae82} Anders E. \& Ebihara M. 1982, Geochimica et Cosmochimica Acta 46, 2363

\bibitem[Arnaud(1996)]{a96} Arnaud, K. A. 1996, ASP Conf. Ser. 101: Astronomical Data Analysis Software and Systems V, ed. Jacoby G. \& Barnes J., 17

\bibitem[Bohlin, Savage \& Drake (1978)]{bsd78} Bohlin, R. C., Savage, B. D., \& Drake, J. F. 1978, \apj, 224, 132

\bibitem[Canizares \etal(2005)]{ca05} Canizares, C. R., \etal 2005, PASP, accepted (astro-ph/0507035)

\bibitem[Cappi \etal(1997)]{ca97} Cappi, M., Matsuoka, M., Comastri, A., Brinkmann, W., Elvis, M., Palumbo, G. G. C., \& Vignali, C. 1997, ApJ, 478, 492

\bibitem[Carilli, Rupen \& Yanny (1993)]{cry93}Carilli, C. L., Rupen, M. P., \& Yanny, Brian 1993, \apj, 412, L59

\bibitem[Chartas \etal(2002)]{ch02} Chartas, G., Agol, E., Eracleous, M., Garmire, G. P., Bautz, M. W., \& Morgan, N. D. 2002, \apj, 568, 509

\bibitem[Dai \etal(2003)]{da03} Dai, X., Chartas, G., Agol, E., Bautz, M. W., \& Garmire, G. P. 2003, \apj, 589, 100

\bibitem[Dai \etal(2004)]{da04} Dai, X., Chartas, G., Eracleous, M., \& Garmire, G. P. 2004, \apj, 605, 45

\bibitem[Dai \& Kochanek (2005)]{dk05} Dai, Xinyu \& Kochanek, Christopher S. 2005, \apj, 625, 633

\bibitem[Dwek (1998)]{dw98} Dwek, Eli 1998, \apj, 501, 643

\bibitem[de Rosa \etal(2005)]{de05} de Rosa, A., Piro, L., Tramacere, A., Massaro, E., Walter, R., Bassani, L., Malizia, A., Bird, A. J., \& Dean, A. J. 2005, \aap, 438, 121

\bibitem[Dickey \& Lockman (1990)]{d90} Dickey, J. M. \& Lockman F. J. 1990, ARA\&A 28, 215

\bibitem[Donato \etal(2001)]{do01} Donato, D., Ghisellini, G., Tagliaferri, G., \& Fossati, G. 2001, \aap, 375, 739

\bibitem[Draine (2003)]{dr03} Draine, B. T. 2003, \araa, 41, 241

\bibitem[Edmunds(2001)]{ed01} Edmunds, M. G. 2001, \mnras, 328, 223

\bibitem[Falco \etal(1999)]{fal99} Falco, E. E., Impey, C. D., Kochanek, C. S., Leh\'ar, J., McLeod, B. A., Rix, H.-W., Keeton, C. R., Mu\~noz, J. A., \& Peng, C. Y. 1999, \apj, 523, 617

\bibitem[Garmire \etal(2003)]{g03} Garmire, G. P., Bautz, M. W., Nousek, J. A., \& Ricker, G. R. 2003, SPIE, 4851, 28

\bibitem[Gerin \etal(1997)]{ge97} Gerin, M., Phillips, T. G., Benford, D. J., Young, K. H.,Menten, K. M., \& Frye, B. 1997, ApJ, 488, L31

\bibitem[G\'omez-\'Alvarez \etal(2004)]{go04} G\'omez-\'Alvarez, P., Mediavilla Gradolph, E., S\'anchez, S. F., Arribas, S., Wisotzki, L., Wambsganss, J., Lewis, G., \& Mu\~noz, J. A. 2004, AN, 325, 132

\bibitem[Hewitt \etal(1992)]{he92}  Hewitt, J. N., Turner, E. L., Lawrence, C. R., Schneider, D. P., \& Brody, J. P. 1992, \aj, 104, 968

\bibitem[Inoue(2003)]{in03} Inoue, A. K. 2003, \pasj, 55, 901

\bibitem[Jauncey \etal(1991)]{ja91} Jauncey, D. L., et al. 1991, Nature, 352, 132

\bibitem[Lawrence \etal(1995)]{la95} Lawrence, C. R., Elston, Richard, Januzzi, B. T., \& Turner, E. L. 1995, \aj, 110, 2570

\bibitem[Lidman \etal(2000)]{li00} Lidman, C., Courbin, F., Kneib, J.-P., Golse, G., Castander, F., \& Soucail, G. 2000, \aap, 364, L62

\bibitem[Lidman \etal(1999)]{li99} Lidman, C., Courbin, F., Meylan, G., Broadhurst, T., Frye, B., \& Welch, W. J. W. 1999, ApJ, 514, L57

\bibitem[Maiolino \etal(2001)]{ma01} Maiolino, R., Marconi, A., Salvati, M., Risaliti, G., Severgnini, P., Oliva, E., La Franca, F., \& Vanzi, L. 2001, \aap, 365, 28

\bibitem[Maiolino, Marconi \& Oliva (2001)]{mmo01} Maiolino, R., Marconi, A., \& Oliva, E. 2001, \aap, 365, 37

\bibitem[Mathur \& Nair (1997)]{mn97} Mathur, Smita \& Nair, Sunita 1997, \apj, 484, 140

\bibitem[McGough \etal(2005)]{mc05} McGough, Christina, Clayton, Geoffrey C., Gordon, Karl D., \& Wolff, Michael J. 2005, \apj, 624, 118

\bibitem[Mediavilla \etal(2005)]{me05} Mediavilla, E., Mu\~noz, J. A., Kochanek, C. S., Falco, E. E., Arribas, S., \& Motta, V. 2005, \apj, 619, 749

\bibitem[Mori \etal(2001)]{m01} Mori, K., Tsunemi, H., Miyata, E., Baluta, C., Burrows, D. N., Garmire, G. P., \& Chartas, G. 2001, in ASP Conf. Ser. 251, New Century of X-Ray Astronomy, ed. H. Inoue \& H. Kunieda (San Francisco: ASP), 576

\bibitem[Morrison \& McCammon (1983)]{mm83} Morrison, R. \& McCammon, D. 1983, \apj, 270, 119

\bibitem[Motta \etal(2002)]{mo02} Motta, V., Mediavilla, E., Mu\~noz, J. A., Falco, E., Kochanek, C. S., Arribas, S., Garc\'{\i}a-Lorenzo, B., Oscoz, A., \& Serra-Ricart, M. 2002, \apj, 574, 719

\bibitem[Mu\~noz \etal(2004)]{mu04} Mu\~noz, J. A., Falco, E. E., Kochanek, C. S., McLeod, B. A., \& Mediavilla, E. 2004, \apj, 605, 614

\bibitem[Nadeau \etal(1991)]{na91} Nadeau, D., Yee, H. K. C., Forrest, W. J., Garnett, J. D., Ninkov, Z., \& Pipher, J. L. 1991, \apj, 376, 430

\bibitem[Pramesh Rao \& Subrahmanyan (1988)]{ps88} Pramesh Rao, A., \& Subrahmanyan, R. 1988, MNRAS, 231, 229

\bibitem[Reeves \& Turner (2000)]{rt00} Reeves, J. N. \& Turner, M. J. L. 2000, \mnras, 316, 234

\bibitem[Subrahmanyan \etal(1990)]{su90} Subrahmanyan, R., Narasimha, D., Pramesh-Rao, A., \& Swarup, G. 1990, MNRAS, 246, 263

\bibitem[Toft, Hjorth \& Burud (2000)]{thb00} Toft, S., Hjorth, J., \& Burud, I. 2000, \aap, 357, 115 

\bibitem[Tonry \& Kochanek (1999)]{tk99}  Tonry, John L. \& Kochanek, Christopher S. 1999, \aj, 117, 2034

\bibitem[Tsunemi \etal(2001)]{t01} Tsunemi, H., Mori, K., Miyata, E., Baluta, C., Burrows, D. N., Garmire, G. P., \& Chartas, G. 2001, \apj, 554, 496

\bibitem[Weisskopf \etal(2002)]{we02} Weisskopf, M. C., Brinkman, B., Canizares, C., Garmire, G., Murray, S., \& Van Speybroeck, L. P. 2002, \pasp, 114, 1 

\bibitem[Wiklind \& Combes (1995)]{wc95}Wiklind, T. \& Combes, F. 1995, \aap, 299, 382

\bibitem[Wiklind \&  Combes (1996)]{wc96}Wiklind, T. \& Combes, F. 1996, Nature, 379, 139

\bibitem[Wiklind \& Combes (1998)]{wc98} Wiklind, T. \& Combes, F. 1998, \apj, 500, 129

\bibitem[Winn \etal(2002)]{wi02} Winn, Joshua N., Kochanek, Christopher S., McLeod, Brian A., Falco, Emilio E., Impey, Christopher D., \& Rix, Hans-Walter 2002, \apj, 575, 103

\bibitem[Wisotzki \etal (1995)]{wi95} Wisotzki, L., Koehler, T., Ikonomou, M., \& Reimers, D. 1995, \aap, 297, 59L

\bibitem[Wisotzki \etal(1993)]{wi93} Wisotzki, L., Koehler, T., Kayser, R., \& Reimers, D. 1993, \aap, 278, L15

\bibitem[Wucknitz \etal(2003)]{wu03} Wucknitz, O., Wisotzki, L., Lopez, S., \& Gregg, M. D. 2003, \aap, 405, 445
\end{thebibliography}
\end{document}